\documentclass[pra,twocolumn,amsmath,amssymb,superscriptaddress]{revtex4-1}
\usepackage{epsfig,amsmath}
\usepackage{subfigure}
\usepackage{graphicx}
\usepackage{dcolumn}
\usepackage{stmaryrd}
\usepackage{mathrsfs}
\usepackage{pifont}
\usepackage{amsthm}
\usepackage{amssymb}
\usepackage{bm}
\usepackage{latexsym}
\usepackage[colorlinks=true,linkcolor=blue,citecolor=blue]{hyperref}
\usepackage{color}
\usepackage{epstopdf}

\usepackage{booktabs}
\usepackage{threeparttable}
\usepackage{multirow}
\usepackage{array}
\usepackage{mathdots}

\begin{document}

\title{Deterministic generation of cat states exceeding 100 photons under dissipation}

\author{Zhu-yao Jin}
\affiliation{School of Physics, Zhejiang University, Hangzhou 310027, Zhejiang, China}

\author{Jun Jing}
\email{Contact author: jingjun@zju.edu.cn}
\affiliation{School of Physics, Zhejiang University, Hangzhou 310027, Zhejiang, China}

\date{\today}

\begin{abstract}
Large-size cat states are of fundamental significance in exploring the quantum-to-classical transition, and serve as promising resources for quantum metrology and fault-tolerant quantum computation. However, scaling up the amplitude of cat states remains challenging because of the growing susceptibility to decoherence and resource overhead in complex operation. We propose to generate large cat states by using the dynamical invariant of hybrid qubit-resonator systems under either a Hermitian or non-Hermitian time-dependent Hamiltonian. This is a study in the framework of universal quantum control (UQC), where the system dynamics is analyzed in the ancillary picture via a unitary transformation conditioned on the qubit state. The controlled dynamics of the composite system can be encoded in the evolution of the dynamical invariant via the Heisenberg equation, which yields constraints on the Hamiltonian. When the qubit is prepared in a balanced superposition state, the bosonic mode can be deterministically prepared from the vacuum state to the cat state of a mean photon number over $120$, following a trade-off between the attainable amplitude, the protocol duration, and the experimental parameters. In the Hermitian case, the generation is perfect. In the non-Hermitian case, the generation is conditional on the postselection of trajectories ignoring quantum jumps associated with the qubit, with a target-state fidelity over $0.95$. Our protocol can also be applied to the generation of large intrinsic cat states and four-legged cat states. Through the preparation of macroscopic quantum states, our work substantially advances UQC to hybrid discrete-continuous variable systems.
\end{abstract}

\maketitle

\section{Introduction}

The intrinsic cat states traces back to Schr\"odinger's gedanken experiment~\cite{Schrodinginger1935Die}, where the cat's two macroscopically distinguished states, i.e., dead and alive, are respectively entangled with the radioactive decay and the subsequent breaking of the poison bottle and the otherwise stable state. This paradigmatic model embarks the investigation of the macroscopic quantum states~\cite{Frowis2018Macroscopic}, expected to address the open question on the boundary between classical and quantum worlds~\cite{Frowis2018Macroscopic,Wineland2013Nobel,Haroche2013Nobel,Arndt2014Testing}.

In the literature, macroscopic quantum states refer mainly to atomic cat states~\cite{Agarwal1997Atomic,Omran2019Generation,Chao2019Generation,Yang2025Minute} and bosonic cat states~\cite{Schleich1991Nonclassical,Gerry1997Quantum}. The size of the atomic cat states is characterized by the atomic number. However, their preparation in reality is hindered by the systematic errors and the dissipation on individual atoms~\cite{Agarwal1997Atomic,Omran2019Generation,Chao2019Generation,Yang2025Minute}, whose effect grows with the system size. The cat states' fidelities observed in $20$ Rydberg atoms~\cite{Omran2019Generation} and in $20$ superconducting qubits~\cite{Chao2019Generation} are about $0.542$ and $0.525$, respectively. In comparison to atomic systems, bosonic modes provide a more versatile platform for generating cat states as superpositions of coherent states~\cite{Schleich1991Nonclassical,Gerry1997Quantum}: $|\alpha\rangle+|-\alpha\rangle$. The coherent bases with $|\alpha|>2$ are approximately orthogonal to each other and thereby cat state can be regarded as a macroscopic quantum state~\cite{Leonhardt1995Measureing,Leggett2002Testing}. Similar superposition states in the massive systems can test macroscopic quantum phenomena~\cite{Marius2023Schrodinger}. Large-amplitude bosonic cat states constitute a foundation for investigating quantum-to-classical transition~\cite{Schleich1991Nonclassical,Gerry1997Quantum}, high-fidelity quantum teleportation~\cite{Brask2010Hybrid,Lee2013Neardeterministic}, enhanced sensitivity of phase estimation~\cite{Gilchrist2004Schrodinger,Joo2011Quantum,Giovannetti2011Advances,Pezze2018Quantum}, and fault-tolerant quantum computation~\cite{Cochrane1999Macroscopically,Ralph2003Quantum,Lund2008Fault,Mirrahimi2014Dynamically,
Grimm2020Stabilization,Chamberland2022Building,Hastrup2022Alloptical,Ozlem2025Unified}. By virtue of quantum error correction~\cite{Ofek2016Extending}, bit-flip errors in cat qubits~\cite{Mirrahimi2014Dynamically,Grimm2020Stabilization} could be exponentially suppressed with increasing amplitude of the cat states~\cite{Putterman2025Preserving,Lescanne2020Exponential}. Large cat states have been observed in many platforms, including trapped-ion systems~\cite{Monroe1996Schrodinger,
Haljan2005Spin,McDonnell2007Long,Lo2015Spin,Johnson2017Ultrafast,Rojkov2026Stabilization}, circuit QED (cQED) systems~\cite{Mooij1999Josephson,Friedman2000Quantum,Brian2013Deterministically,Grimm2020Stabilization}, and circuit quantum acoustodynamics systems~\cite{Marius2023Schrodinger}, e.g., a cat state with amplitude $|\alpha|\sim\sqrt{7}$ and fidelity $\sim0.81$ has been generated in a cQED system~\cite{Brian2013Deterministically}. And optical cat states of the electromagnetic wave fields were demonstrated with amplitude $|\alpha|\sim1.85$, fidelity $\sim0.99$, and success probability $0.2$~\cite{Sychev2017Enlargement}.

Besides the two-legged cat states, the four-legged cat states or so-called compass states~\cite{Mirrahimi2014Dynamically,Zurek2001SubPlanck,Leghtas2013Deterministic,
Leghtas2013Hardware,Brian2013Deterministically,Su2015Quantum,Zheng2026Quantum,He2026Generation,
Chen2026Fault}: $|\alpha\rangle+|-\alpha\rangle+|i\alpha\rangle+|-i\alpha\rangle$ are more promising candidates for quantum technologies, such as quantum metrology~\cite{Su2015Quantum}, dark matter searching~\cite{Zheng2026Quantum}, and quantum error syndrome tracking and correction in the presence of photon loss~\cite{Leghtas2013Deterministic,Leghtas2013Hardware,Brian2013Deterministically,Sun2014Tracking}. Recently, a four-legged cat state has been demonstrated in a cQED system~\cite{Zheng2026Quantum} with amplitude $|\alpha|\sim\sqrt{12}$ and fidelity $>0.95$, yet the success probability is lower than $0.01$.

Existing theoretical protocols for the cat-state generation are diverse in their own advantages and constraints. The adiabatic protocol is based on slowly tuning the two-photon driving field of a Kerr-nonlinear resonator~\cite{Cirac1998Quantum,Goto2016Universal,Puri2017Engineering}, whose capability is severely limited by the prolonged exposure to the environment. To mitigate this deficiency, the evolution can be accelerated by the shortcuts to adiabaticity (STA)~\cite{Puri2017Engineering,Hatomura2018Shortcuts} via an extra two-photon driving field. The attainable cat-state amplitude is about $|\alpha|<2$, subject to the ratio of ancillary driving strength to the Kerr nonlinearity~\cite{Grimm2020Stabilization,Iyama2023Observation}. Under two-photon driving field~\cite{Chen2021Shortcuts}, STA generates a cat state in the Jaynes-Cummings model with $|\alpha|\sim2$ within an extremely high squeezed frame of a hardly accessible squeezing-level about $\sim 15$ dB. In addition, by combining a quantum gate that transfers an arbitrary qubit state into a superposition of two quasiorthogonal cavity coherent states of opposite phases~\cite{Leghtas2013Deterministic,Leghtas2013Hardware,Brian2013Deterministically,Hacker2019Deterministic} with sequential pulses on qubit and cavity, one can deterministically generate a cat state with amplitude $|\alpha|\sim\sqrt{7}$ and fidelity $\sim0.91$~\cite{Leghtas2013Deterministic}. Using the photon subtraction method~\cite{Lund2004Conditional,
Neergaard2006Generation,Gerrits2010Generation,Laghaout2013Amplification,Sychev2017Enlargement,Takase2021Generation,
Sun2021Remote} on a squeezed vacuum state with a squeezing level above $15$ dB~\cite{Takase2021Generation}, a large cat state was predicted with $|\alpha|\sim\sqrt{10}$ and fidelity $>0.99$~\cite{Takase2021Generation}, yet the protocol is probabilistic with a success probability $\sim0.02$. As for the stabilization of high-fidelity cat states, the two-photon loss rate is required to dominate the single-photon loss rate~\cite{Gilles1994Generation,Leghtas2015Confining,
Zapletal2022Stabilization}. Theoretically and experimentally, it is generally challenging to create large cat states in open quantum systems~\cite{Myatt2000Decoherence,Mirrahimi2014Dynamically}. A unified theoretical framework for both closed and open quantum systems is desired for deterministically generating the large-size cat states with accessible control.

Our universal quantum control (UQC) framework~\cite{Jin2025Universal,Jin2025UniNon,
Jin2026Unibosonic,Jin2026Nonbosonic,Jin2026Liouville} provides a much broader perspective for controlling the dynamics of classical systems~\cite{Jin2026Liouville} and both discrete- and continuous-variable quantum systems under either Hermitian or non-Hermitian Hamiltonian~\cite{Jin2025Universal,Jin2025UniNon,Jin2026Unibosonic,Jin2026Nonbosonic,Jin2026Liouville}. Fundamentally it is based on the dynamical invariants of systems, which are constructed through the gauge potential induced by representation transformation. In this paper, to generate large-amplitude cat states, we construct the dynamical invariants for the hybrid qubit-resonator systems with longitudinal interaction. For the Hermitian systems, a dynamical invariant is constructed to satisfy the Heisenberg equation with the rotated Hamiltonian, giving rise to the constraint conditions on the original Hamiltonian. When the qubit is initially in the balanced superposition state and the bosonic mode is in the vacuum state, the mode can be faithfully prepared as a cat state with mean photon number $\sim120$. In the open-quantum-system scenario, we consider the quantum trajectories without quantum jumps related to qubit. Then the dynamical invariant satisfies the Heisenberg equation with the non-Hermitian Hamiltonian, thereby imposing the constraints associated with the gain and loss rates. Under the conditional probability conservation, the cat states of the same mean photon number as the closed system can be almost deterministically generated with a fidelity $\mathcal{F}>0.962$. Moreover, our protocol applies to the generation of the intrinsic cat states and the four-legged cat states with record high fidelity and amplitudes.

The rest of this paper is structured as follows. The UQC theory is briefly recalled in Secs.~\ref{generalHerm} and \ref{generalNonHerm} for the Hermitian and non-Hermitian systems, respectively, demonstrating controlled system dynamics via dynamical invariants. In Sec.~\ref{SecClose}, we apply the general theory to the closed hybrid qubit-resonator systems and construct the relevant dynamical invariant to generate large cat states. In Sec.~\ref{opennojump}, the general theory is applied to the open-system scenario described by non-Hermitian Hamiltonian. In Sec.~\ref{fullmaster}, we analyze the impacts from the approximations, including the quantum-jump terms in the Lindblad master equation and the abrupt switching of control parameters. Our protocol is extended in Secs.~\ref{SubReal} and \ref{fourcomponent} to the generation of large intrinsic cat states and four-legged cat states, respectively. The entire work is summarized in Sec.~\ref{Conclud}. Appendix~\ref{EffHam} details the derivation for the non-Hermitian Hamiltonian from the Lindblad master equation of the hybrid qubit-resonator systems.

\section{Universal quantum control}\label{general}

This section briefly outlines the universal quantum control theory~\cite{Jin2025Universal,Jin2025UniNon,
Jin2026Unibosonic,Jin2026Nonbosonic,Jin2026Liouville} for the time-dependent systems, Hermitian or non-Hermitian. This framework substantially demonstrates the controlled system dynamics via the dynamical invariants and the gauge potential. The gauge potential can be introduced by the transformation between the time-dependent and the stationary ancillary representations.

\subsection{Hermitian systems}\label{generalHerm}

The dynamics of a closed quantum system is generally described by the Schr\"odinger equation as ($\hbar\equiv1$)
\begin{equation}\label{Sch}
i\frac{\partial}{\partial t}|\psi(t)\rangle=H(t)|\psi(t)\rangle,
\end{equation}
with the time-dependent Hermitian Hamiltonian $H(t)$ and the pure-state solution $|\psi(t)\rangle$. Although the solutions to Eq.~(\ref{Sch}) can be obtained in principle from the instantaneous eigenvalue equations, such a brute-force approach will become increasingly intractable as the system size grows.

Alternatively, the system dynamics can be described in the ancillary representation via a time-dependent unitary transformation $\mathcal{V}(t)$~\cite{Jin2025Universal,
Jin2025UniNon,Jin2026Unibosonic,Jin2026Nonbosonic,Jin2026Liouville}. In the relevant rotating frame, the system Hamiltonian is transformed as
\begin{equation}\label{Hamrot}
H_{\rm rot}(t)=\mathcal{V}^\dagger(t)H(t)\mathcal{V}(t)
-i\mathcal{V}^\dagger(t)\frac{\partial\mathcal{V}(t)}{\partial t},
\end{equation}
where the second term describes the holonomic contribution in the Hilbert space~\cite{Zhang2023Geometric} or the gauge potential. One can impose constraints on the rotated Hamiltonian~(\ref{Hamrot}) by requiring the stationary dynamical invariants $\mathcal{J}(0)$ to satisfy the Heisenberg equation as~\cite{Jin2026Liouville}:
\begin{equation}\label{DynInva}
\frac{\partial\mathcal{J}(0)}{\partial t}=-i\left[H_{\rm rot}(t),\mathcal{J}(0)\right],
\end{equation}
where $\mathcal{J}(0)=\tilde{\mu}(0)$ or $\tilde{\mu}^\dagger(0)$. The invariant can be accompanied with a real-valued gauge-invariant global phase $f(t)$, e.g., $\tilde{\mu}(0)\equiv\exp[if(t)]\mu(0)$. In the continuous-variable systems~\cite{Jin2026Liouville}, $\mu(0)$ and $\mu^\dagger(0)$ are superposed by the bosonic operators in the original picture.

In accordance with Eq.~(\ref{DynInva}), we can have the time-dependent dynamical invariants $\tilde{\mu}(t)=\mathcal{V}(t)\mathcal{J}(0)\mathcal{V}^\dagger(t)$, which satisfy the Heisenberg equation with the original Hamiltonian~(\ref{Sch}). In particular, under the inversion of Eq.~(\ref{Hamrot}), Eq.~(\ref{DynInva}) is transformed as
\begin{equation}\label{vonNeuTurn}
\begin{aligned}
&\frac{\partial[\mathcal{V}^\dagger(t)\mathcal{J}(t)\mathcal{V}(t)]}{\partial t}=-i\Big[\mathcal{V}^\dagger(t)H(t)\mathcal{V}(t)\\
&-i\mathcal{V}^\dagger(t)\frac{\partial\mathcal{V}(t)}{\partial t}, \mathcal{V}^\dagger(t)\mathcal{J}(t)\mathcal{V}(t)\Big].
\end{aligned}
\end{equation}
By the chain rule, we have
\begin{equation}\label{vonNeuChain}
\begin{aligned}
&\mathcal{V}^\dagger(t)\frac{\partial\mathcal{J}(t)}{\partial t}\mathcal{V}(t)=-i\mathcal{V}^\dagger(t)\left[H(t), \mathcal{J}(t)\right]\mathcal{V}(t)\\
&+\mathcal{V}^\dagger(t)\mathcal{J}(t)\frac{\partial\mathcal{V}(t)}{\partial t}-\mathcal{V}^\dagger(t)\frac{\partial\mathcal{V}(t)}{\partial t}\mathcal{V}^\dagger(t)\mathcal{J}(t)\mathcal{V}(t)\\
&-\frac{\partial\mathcal{V}^\dagger(t)}{\partial t}\mathcal{J}(t)\mathcal{V}(t)
-\mathcal{V}^\dagger(t)\mathcal{J}(t)\frac{\partial\mathcal{V}(t)}{\partial t}.
\end{aligned}
\end{equation}
By left-multiplying Eq.~(\ref{vonNeuChain}) with $\mathcal{V}(t)$ and right-multiplying it with $\mathcal{V}^\dagger(t)$, and using the relation $d[\mathcal{V}(t)\mathcal{V}^\dagger(t)]/dt=[d\mathcal{V}(t)/dt]\mathcal{V}^\dagger(t)
+\mathcal{V}(t)[d\mathcal{V}^\dagger(t)/dt]=0$, it ends up with
\begin{equation}\label{DynInvaOri}
\frac{\partial\mathcal{J}(t)}{\partial t}=-i[H(t), \mathcal{J}(t)].
\end{equation}
This self-consistent proof confirms that under the constraints imposed on $H(t)$ by Eq.~(\ref{DynInvaOri}), the system dynamics can be described by either evolution of the two conjugate dynamical invariants:
\begin{equation}\label{evolveHerm}
\mu(0)\rightarrow e^{if(t)}\mu(t)\quad \& \quad  \mu^\dagger(0)\rightarrow e^{-if(t)}\mu^\dagger(t).
\end{equation}
In various formulations, the dynamical invariants render ubiquitous applications, covering both discrete-~\cite{Jin2025Universal} and continuous-variable systems~\cite{Jin2026Unibosonic}.

\subsection{Non-Hermitian systems}\label{generalNonHerm}

Realistic systems are invariably open: gain and loss generically break the self-adjointness of the system Hamiltonian. Nevertheless, our theory based on dynamical invariants and gauge potential also enables controlled dynamics of the open quantum systems~\cite{Jin2025UniNon,Jin2026Nonbosonic,Jin2026Liouville}. For discrete-variable systems, the dynamics can be described by the non-Hermitian Hamiltonian that is directly derived from Lindblad master equation by ignoring the quantum jump terms under postselection~\cite{Han2024Measuring}. In contrast, for the continuous-variable systems governed by the quadratic Hamiltonian, the non-Hermitian Hamiltonian can be rigorously derived from the adjoint Lindblad master equation with all the jump terms retained~\cite{Metelmann2015Nonreciprocal,Wang2019Nonreciprocity,Jin2026Nonbosonic,Jin2026Liouville}. Under the biorthogonal assumption~\cite{Brody2013Biorhogonal}, the dynamics of non-Hermitian systems can be described by two sets of time-dependent Schr\"odinger equation as
\begin{equation}\label{SchNon}
i\frac{\partial}{\partial t}|\psi(t)\rangle=H(t)|\psi(t)\rangle,\quad i\frac{\partial}{\partial t}|\phi(t)\rangle=H^\dagger(t)|\phi(t)\rangle,
\end{equation}
with $H(t)\neq H^\dagger(t)$, and $|\psi(t)\rangle$ and $\langle\phi(t)|$ are the pure-state solutions in the ket and bra spaces, respectively.

Without loss of generality, we focus on the system dynamics in the ket space governed by $H(t)$ in the following. Similar to the Hermitian case in Eq.~(\ref{Hamrot}), in the rotating frame with respect to the unitary transformation $\mathcal{V}(t)$~\cite{Jin2026Nonbosonic,Jin2026Liouville}, we have a transformed Hamiltonian $H_{\rm rot}(t)$. The dynamical component $\mathcal{V}^\dagger(t)H(t)\mathcal{V}(t)$ is non-Hermitian and the geometrical component $-i\mathcal{V}^\dagger\dot{\mathcal{V}}$ remains Hermitian. Thus the dynamical invariant $\mu(t)$ and its Hermitian conjugate $\mu^\dagger(t)$ cannot simultaneously satisfy the Heisenberg equation~(\ref{DynInvaOri}) under the same constrained Hamiltonian. In other words, only one of $\mu(t)$ and $\mu^\dagger(t)$ can be activated as the universal passages for controlling the system dynamics~\cite{Jin2026Nonbosonic,Jin2026Liouville}. Accordingly, the system dynamics is effectively demonstrated as
\begin{equation}\label{evolveNonHerm}
\mu(0)\rightarrow e^{if(t)}\mu(t)\quad |\quad \mu^\dagger(0)\rightarrow e^{-if(t)}\mu^\dagger(t),
\end{equation}
where $f(t)$ becomes a complex-valued global phase, in sharp contrast to Eq.~(\ref{evolveHerm}). The imaginary part of $f(t)$ encodes gain or loss induced by the system-environment interaction, leading to the probability nonconservation of the system. By a proper choice of systematic parameters, $f(t)$ could become a real number at a predesigned moment $\tau$, corresponding to the automatic normalization of the system wave-function. Also, the non-Hermitian dynamical invariants have enabled perfect transfer of arbitrary states in both discrete-~\cite{Jin2025UniNon} and continuous-variable systems~\cite{Jin2026Nonbosonic} and the generation of ultra highly squeezed states in open quantum systems~\cite{Jin2026Liouville}. Next, we extend the UQC theory~\cite{Jin2026Nonbosonic,Jin2026Liouville} to construct the dynamical invariants for a hybrid discrete-continuous variable system.

\section{Generation of cat state in closed qubit-resonator system}\label{SecClose}

\begin{figure}[htbp]
\centering
\includegraphics[width=0.8\linewidth]{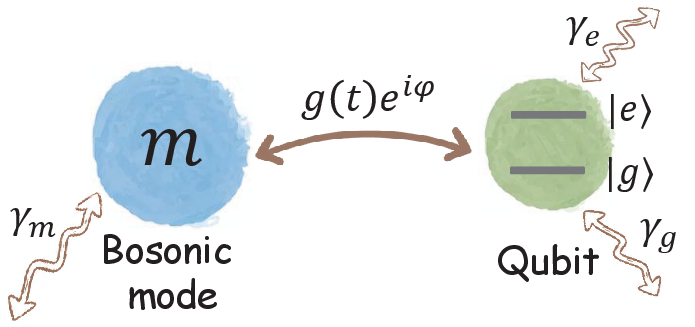}
\caption{Sketch of the open hybrid qubit-bosonic mode system, where the qubit is longitudinally coupled to the bosonic mode with the strength $g(t)$ and the phase $\varphi$. $\gamma_e$, $\gamma_g$, and $\gamma_m$ denote the gain or loss rates of the levels $|e\rangle$, $|g\rangle$, and the bosonic mode, respectively. When $\gamma_a=0$, $a=e,g,m$, it becomes a closed system.}\label{model}
\end{figure}

Our study in this section is conducted on a closed hybrid quantum system consisting of a qubit and a bosonic mode as shown in Fig.~\ref{model} with vanishing gain or loss rates $\gamma_a=0$, $a=e,g,m$. The qubit is composed of an excited state $|e\rangle$ and a ground state $|g\rangle$, which is coupled to the bosonic mode by the longitudinal interaction with strength $g(t)$, modulation frequency $\omega_0$, and modulation phase $\varphi(t)$. The Hamiltonian of the composite system~\cite{Didier2015Fast,Blais2021Circuit} can be expressed as
\begin{equation}\label{Ham}
\begin{aligned}
H(t)&=\omega_m(t)m^\dagger m+\frac{\omega_q(t)}{2}\sigma_z\\ &+2g(t)\cos(\omega_0t+\varphi(t))\sigma_z\left(m+m^\dagger\right),
\end{aligned}
\end{equation}
where $m$ ($m^\dagger$) denotes the annihilation (creation) operator of the bosonic mode, and the Pauli operator is defined as $\sigma_z\equiv|e\rangle\langle e|-|g\rangle\langle g|$. $\omega_m(t)$ and $\omega_q(t)$ are time-modulated frequencies of the bosonic mode and qubit, respectively. In the rotating frame with respect to $H_0=\omega_0m^\dagger m$, and under the rotating-wave approximation assuming $\omega_0\gg g(t)$, we have
\begin{equation}\label{Hamprime}
\begin{aligned}
H(t)&=\Delta_m(t)m^\dagger m+\frac{\omega_q(t)}{2}\sigma_z\\
&+g(t)\sigma_z\left(me^{i\varphi(t)}+m^\dagger e^{-i\varphi(t)}\right),
\end{aligned}
\end{equation}
where $\Delta_m(t)=\omega_m(t)-\omega_0$ denotes the detuning between the bosonic-mode frequency and the modulation frequency. The Hamiltonians~(\ref{Ham}) and (\ref{Hamprime}) are available in cQED systems~\cite{Blais2004Cavity,Schuster2005ac,Wallraff2005Approaching,Brian2013Deterministically,Reagor2016Quantum,
Wang2017Converting,Ian2025Hot} and trapped-ion systems~\cite{Katz2023Programmable,Lee2019Ion}. The longitudinal interaction $g(t)$ between transmon qubit and LC oscillator can be tuned by the external flux~\cite{Billangeon2015Circuit,Didier2015Fast,Richer2016Circuit,Royer2017fasthighfidelity,Blais2021Circuit}, and that between the electron spin and microwave cavity is tunable via the external magnetic field~\cite{Harvey2018Coupling,Bottcher2022Parametric,Bosco2022Fully}. For the bosonic mode with a coherence time $T_2\sim1\mu$s, the longitudinal interaction $g(t)/2\pi$ typically lies in the range of $[10, 10^3]$ MHz~\cite{Didier2015Fast,Bosco2022Fully}.

A proper representation is desired to construct the dynamical invariant to generate cat states. We suppose that the relevant unitary transformation $\mathcal{V}(t)$ in Eq.~(\ref{Hamrot}) takes a conditional form as
\begin{equation}\label{Vspin}
\mathcal{V}(t)=D^\dagger[\alpha(t)]\otimes|e\rangle\langle e|+D[\alpha(t)]\otimes|g\rangle\langle g|,
\end{equation}
where $D(\alpha)\equiv\exp(\alpha m^\dagger-\alpha^*m)$ is a displacement operator with $\alpha(t)=\theta(t)\exp[-i\beta(t)]$. Here $\theta(t)$ and $\beta(t)$ are real and time-dependent amplitude and phase, respectively. In the rotating frame with respect to $\mathcal{V}(t)$ in Eq.~(\ref{Vspin}), $H(t)$ in Eq.~(\ref{Hamprime}) is transformed as
\begin{equation}\label{Hamrotspin}
\begin{aligned}
&H_{\rm rot}(t)=\mathcal{V}^\dagger(t)H(t)\mathcal{V}(t)
-i\mathcal{V}^\dagger(t)\frac{\partial\mathcal{V}(t)}{\partial t}\\
=&[H_e(t)-\mathcal{A}_e(t)]\otimes|e\rangle\langle e|+[H_g(t)-\mathcal{A}_g(t)]\otimes|g\rangle\langle g|.
\end{aligned}
\end{equation}
Here we have two conditional dynamical components:
\begin{equation}\label{DynCondi}
\begin{aligned}
&H_{e,g}(t)=\Delta_mm^\dagger m\pm\frac{\omega_q}{2}-\Delta_m\theta^2(t)\\
\pm&\left\{\left[g(t)e^{i\varphi(t)}-\Delta_m\theta(t)e^{i\beta(t)}\right]\left[m\mp\theta(t)e^{-i\beta(t)}\right]+{\rm H.c.}\right\},
\end{aligned}
\end{equation}
and two conditional gauge potentials:
\begin{equation}\label{ACondi}
\begin{aligned}
\mathcal{A}_{e,g}(t)&=\pm i\Big\{\partial_t\left[\theta(t)e^{i\beta(t)}\right]\left[m\mp\theta(t)e^{-i\beta(t)}\right]\\
&-\partial_t\left[\theta(t)e^{-i\beta(t)}\right]\left[m^\dagger\mp\theta(t)e^{i\beta(t)}\right]\Big\},
\end{aligned}
\end{equation}

\begin{figure*}[htbp]
\centering
\includegraphics[width=0.95\linewidth]{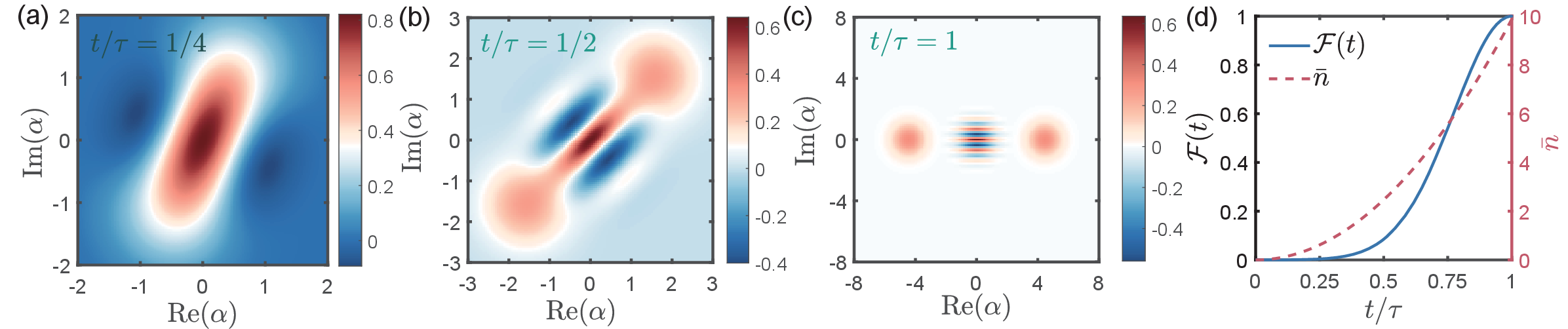}
\includegraphics[width=0.95\linewidth]{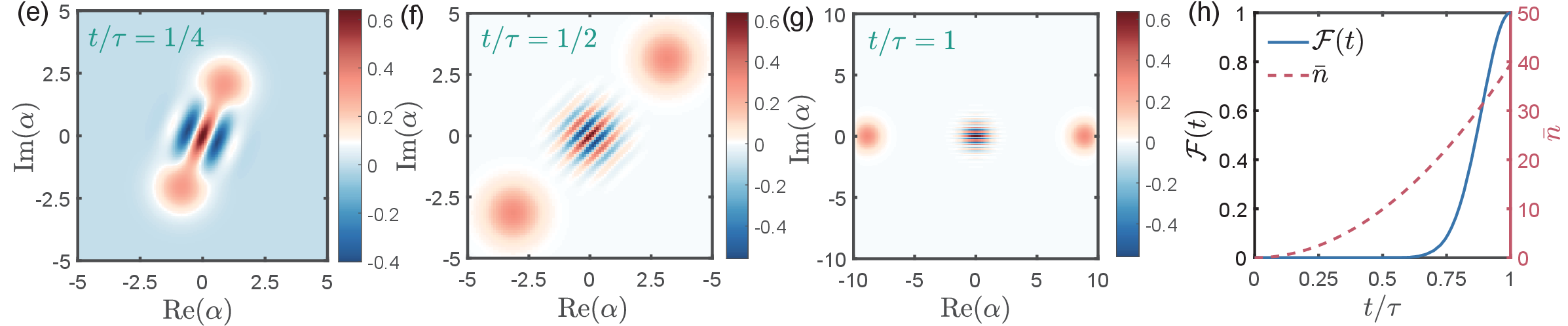}
\includegraphics[width=0.95\linewidth]{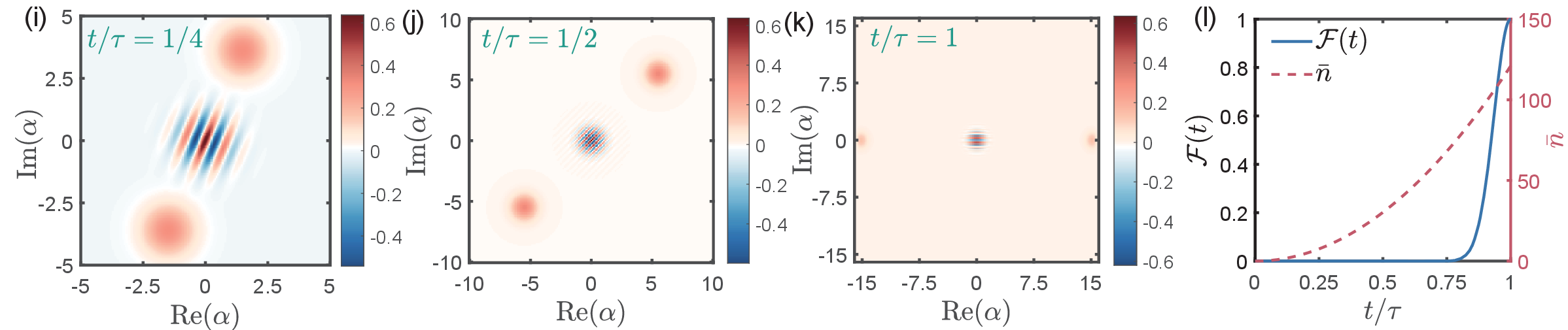}
\caption{Wigner functions of specific moments $t/\tau$ describing the evolution of the resonator from vacuum state to cat state. The first three panels of each line, e.g., (a-c), (e-g), and (i-k) describe the instantaneous resonator states $|\psi_b(t)\rangle$ with various amplitudes $\theta$ or $|\alpha|$, obtained upon the projective measurement on the qubit basis $|+\rangle$. The relevant dynamics about the target state fidelity and the mean photon number for the final entangled states $|\Psi(\tau)\rangle=(|\mathcal{C}_\alpha^+(\tau)\rangle\otimes|+\rangle
-|\mathcal{C}_\alpha^-(\tau)\rangle\otimes|-\rangle)/\sqrt{2}$ appears at the rightmost panel of each line, e.g., (d), (h), and (l). The protocol period is set as $\tau\sim1\mu$s. The system is initially in the product state $|\psi(0)\rangle=|0\rangle\otimes|+\rangle$. The detuning of the bosonic mode $\Delta_m(t)$, the longitudinal interaction $g(t)$, and the phase $\varphi(t)$ are set by Eq.~(\ref{Constr}) with $\beta(t)=\pi t/(2\tau)+\pi/2$ and in (a-d) $\theta(t)=\pi t/\tau$, in (e-h) $\theta(t)=2\pi t/\tau$, and in (i-l) $\theta(t)=7\pi t/2\tau$. }\label{Hermdyn}
\end{figure*}

To cancel the linear terms of $H_{\rm rot}(t)$ in Eq.~(\ref{Hamrotspin}), one can impose constraints on detuning, coupling strength, and coupling phase in Eq.~(\ref{DynCondi}), in accordance with the gauge potentials in Eq.~(\ref{ACondi}). In particular, we have
\begin{equation}\label{Constr}
\Delta_m(t)=\dot{\beta}(t),\quad g(t)=\dot{\theta}(t),\quad \varphi(t)=\beta(t)+\frac{\pi}{2}.
\end{equation}
And we have no constraint on the qubit eigenfrequency. Equation~(\ref{Constr}) suggests that the systematic parameters are not necessarily time-dependent, provided that $\theta(t)$ and $\beta(t)$ can be chosen as a linear function of time and constant, respectively. Using Eq.~(\ref{Constr}), $H_{\rm rot}(t)$ in Eq.~(\ref{Hamrotspin}) is reduced to
\begin{equation}\label{HamrotRed}
H_{\rm rot}(t)=\Delta_m(t)m^\dagger m+\frac{\omega_q(t)}{2}\sigma_z-\Delta_m(t)\theta^2(t).
\end{equation}
Then according to Eq.~(\ref{DynInva}), the relevant dynamical invariant with a vanishing global phase is found to be
\begin{equation}\label{Invaspin}
\mu(0)=|0\rangle\langle0|\otimes|e\rangle\langle e|+|0\rangle\langle 0|\otimes|g\rangle\langle g|,
\end{equation}
where $|0\rangle$ is the vacuum state of the bosonic mode. Using $\mathcal{V}(t)$ in Eq.~(\ref{Vspin}), the evolution of $\mu(0)$ in the original picture can be obtained as
\begin{equation}\label{InvaspinOri}
\begin{aligned}
&\mu(t)=\mathcal{V}(t)\mu(0)\mathcal{V}^\dagger(t)\\
&=|-\alpha(t)\rangle\langle-\alpha(t)|\otimes|e\rangle\langle e|
+|\alpha(t)\rangle\langle\alpha(t)|\otimes|g\rangle\langle g|.
\end{aligned}
\end{equation}
This result indicates that when the composite system is initially in the product state $|\psi(0)\rangle=|0\rangle\otimes|\pm\rangle$ with $|\pm\rangle\equiv(|e\rangle\pm|g\rangle)/\sqrt{2}$, it will become a highly entangled state $|\Psi(\tau)\rangle=(|\mathcal{C}^+_\alpha(\tau)\rangle
\otimes|\pm\rangle\mp|\mathcal{C}^-_\alpha(\tau)\rangle\otimes|\mp\rangle)/\sqrt{2}$ at the desired moment $\tau$, where the cat states are defined as $|\mathcal{C}^\pm_\alpha(t)\rangle\equiv[|\alpha(t)\rangle\pm|-\alpha(t)\rangle]/\mathcal{N}$ with the normalized coefficient $\mathcal{N}$. Upon a projective measurement on the qubit basis $|\pm\rangle$ or $|\mp\rangle$ at $t=\tau$, the bosonic mode will be prepared as the cat state $|\mathcal{C}^+_\alpha(\tau)\rangle$ or $|\mathcal{C}^-_\alpha(\tau)\rangle$ with a balanced probability of $50\%$.

With no loss of generality, the composite system is assumed to be initialized as a product state $|\psi(0)\rangle=|0\rangle\otimes|+\rangle$. The system wave-function $|\psi(t)\rangle$ is numerically obtained by the Schr\"odinger equation $id|\psi(t)\rangle/dt=H(t)|\psi(t)\rangle$ with the full Hamiltonian~(\ref{Ham}) or (\ref{Hamprime}). The protocol performance can be evaluated by the fidelity $\mathcal{F}(t)=|\langle\Psi(\tau)|\psi(t)\rangle|^2$ with the target state $|\Psi(\tau)\rangle=(|\mathcal{C}^+_\alpha(\tau)\rangle\otimes|+\rangle
-|\mathcal{C}^-_\alpha(\tau)\rangle\otimes|-\rangle)/\sqrt{2}$ at the target moment $\tau$ and the mean photon number $\bar{n}\equiv\langle\psi(t)|m^\dagger m|\psi(t)\rangle$. Upon a projection $|+\rangle\langle+|$, the composite system collapses to the tensor product of a normalized wave function for the resonator $|\psi_b(t)\rangle=\sqrt{2}\langle+|\psi(t)\rangle$ and the qubit state $|+\rangle$. The former can be described by the Wigner function in phase space $W(\alpha)\equiv(2/\pi){\rm Tr}[\rho_b(t)D(\alpha)(-1)^{m^\dagger m}D^\dagger(\alpha)]$ with the bosonic density matrix $\rho_b(t)=|\psi_b(t)\rangle\langle\psi_b(t)|$. Note $\bar{n}=\langle\psi_b(t)|m^\dagger m|\psi_b(t)\rangle$.

Figure~\ref{Hermdyn} demonstrates the Wigner function about the resonator wave-function $|\psi_b(t)\rangle$ at certain moments and the fidelity and mean photon number with respect to various target cat states. Through a precise control over the amplitudes and phases, our protocol generates large-size cat states with unit fidelity. In particular, Figs.~\ref{Hermdyn}(a), (b), and (c) present the growing size and the rotation in phase space of the resonator states at $t=\tau/4$, $\tau/2$, and $\tau$, which are characterized by the amplitudes and the phases $(\theta, \beta)=(\pi/4, 5\pi/8)$, $(\pi/2, 3\pi/4)$, and $(\pi, \pi)$, respectively. Figure~\ref{Hermdyn}(d) confirms the quality of the final highly entangled cat state with fidelity $\mathcal{F}(\tau)=1$ and the mean photon number $\bar{n}\sim10$ at $t=\tau$. Similarly, the panels of the second line of Fig.~\ref{Hermdyn} present the generation process toward the cat state with amplitude $|\alpha(\tau)|=\theta(\tau)=2\pi$. By comparing Fig.~\ref{Hermdyn}(g) and Fig.~\ref{Hermdyn}(c), a larger cat state can be verified with a longer distance between the two peaks in its Wigner function. And in the end of the protocol, $t=\tau$, a large cat state is perfectly generated with $\bar{n}\sim40$, as shown in Fig.~\ref{Hermdyn}(h). Comparing the second line to the first line, it is found that the interference pattern becomes clearer with a greater change rate of $\theta(t)$. The third line of Fig.~\ref{Hermdyn} demonstrates the generation of an ultra-large cat state with amplitude $|\alpha|=\theta>10$. As shown in Fig.~\ref{Hermdyn}(l), the target highly entangled state with $\theta(\tau)=3.5\pi$ is confirmed with $\mathcal{F}(\tau)=1$ and $\bar{n}\sim120$. Note the final size of the cat state is only subject to the accessible duration $\tau$ and the coherent time of the composite system in practice, but not limited by the protocol itself.

\section{Generation of cat state in open qubit-resonator system}\label{Secopen}

\subsection{non-Hermitian Hamiltonian under partial postselection}\label{opennojump}

In this section, our protocol of the generation of large cat states is extended to the open quantum systems. The dynamics of the open hybrid qubit-resonator system drafted in Fig.~\ref{model} is generally described by the Lindblad master equation but can be effectively described by a non-Hermitian Hamiltonian under postselection associated with the qubit. Appendix~\ref{EffHam} provides the derivation details and the full dynamics in the presence of all the quantum jump terms is left to Sec.~\ref{fullmaster}. The non-Hermitian Hamiltonian can be expressed as
\begin{equation}\label{HamNon}
\begin{aligned}
H(t)=&\left[\Delta_m(t)-i\frac{\gamma_me^{i\varphi_m}}{2}\right]m^\dagger m\\ +&\frac{1}{2}\left[\omega_q(t)\sigma_z
-\frac{i}{2}\left(\gamma_ee^{i\varphi_e}|e\rangle\langle e|+\gamma_ge^{i\varphi_g}|g\rangle\langle g|\right)\right]\\ +&g(t)\sigma_z\left(me^{i\varphi(t)}+m^\dagger e^{-i\varphi(t)}\right),
\end{aligned}
\end{equation}
where $\gamma_m>0$, $\gamma_e>0$, and $\gamma_g>0$ denote the gain or loss rates of the bosonic mode and the levels $|e\rangle$ and $|g\rangle$, respectively, and the effects of the gain or loss depend on the phases $\varphi_m$, $\varphi_e$, and $\varphi_g$. Without loss of generality, we assume $\gamma_e=\gamma_g=\gamma$, $\varphi_e=\varphi_q$ and $\varphi_g=\varphi_q+\pi$ as follows. Then the non-Hermitian Hamiltonian~(\ref{HamNon}) is reduced to
\begin{equation}\label{HamNonRed}
\begin{aligned}
H(t)&=\tilde{\Delta}_m(t)m^\dagger m+\frac{\tilde{\omega}_q(t)}{2}\sigma_z\\
&+g(t)\sigma_z\left(me^{i\varphi(t)}+m^\dagger e^{-i\varphi(t)}\right),
\end{aligned}
\end{equation}
with $\tilde{\Delta}_m(t)=\Delta_m(t)-i\gamma_m\exp(i\varphi_m)/2$ and $\tilde{\omega}_q(t)=\omega_q(t)-i\gamma\exp{(i\varphi_q)}/2$. If $\varphi_m=\varphi_q=\pi$ or $\varphi_m=\varphi_q=0$, then the composite system is under gain or loss, respectively.

In the non-Hermitian scenario, the conditional unitary transformation $\mathcal{V}(t)$ in Eq.~(\ref{Vspin}) has to be modified with extra amplification or attenuation in the displaced amplitudes. Then we have
\begin{equation}\label{DispNon}
\tilde{\mathcal{V}}(t)=D^\dagger[\tilde{\alpha}(t)]\otimes|e\rangle\langle e|+D[\tilde{\alpha}(t)]\otimes|g\rangle\langle g|,
\end{equation}
where $D[\tilde{\alpha}]$ is the time-dependent displacement operator with $\tilde{\alpha}(t)=\theta(t)\exp[-i\beta(t)-\beta_i(t)]$. $\theta(t)$, $\beta(t)$, and $\beta_i(t)$ are time-dependent real parameters. In the rotating frame with respect to $\tilde{\mathcal{V}}(t)$ in Eq.~(\ref{DispNon}), $H(t)$ in Eq.~(\ref{HamNonRed}) is transformed as
\begin{equation}\label{HamrotNon}
\begin{aligned}
&H_{\rm rot}(t)=\tilde{\mathcal{V}}^\dagger(t)H(t)\tilde{\mathcal{V}}(t)
-i\tilde{\mathcal{V}}^\dagger(t)\frac{\partial\tilde{\mathcal{V}}(t)}{\partial t}\\
=&\left[\tilde{H}_e(t)-\tilde{\mathcal{A}}_e(t)\right]\otimes|e\rangle\langle e|+\left[\tilde{H}_g(t)-\tilde{\mathcal{A}}_g(t)\right]\otimes|g\rangle\langle g|,
\end{aligned}
\end{equation}
where the non-Hermitian conditional dynamical components $\tilde{H}_e(t)$ and $\tilde{H}_g(t)$, and the conditional gauge potentials $\tilde{\mathcal{A}}_e$ and $\tilde{\mathcal{A}}_g$, take the similar forms as those in Eqs.~(\ref{DynCondi}) and (\ref{ACondi}), respectively, under the replacements of $\Delta_m(t)\rightarrow\tilde{\Delta}_m(t)$, $\omega_q(t)\rightarrow\tilde{\omega}_q(t)$, and $\alpha(t)\rightarrow\tilde{\alpha}(t)$.

Using $\tilde{\mathcal{A}}_e$ and $\tilde{\mathcal{A}}_g$, the linear terms of $H_{\rm rot}(t)$ in Eq.~(\ref{HamrotNon}) can be neutralized by imposing the following constraints on the dynamical components:
\begin{equation}\label{ConstrNon}
\begin{aligned}
&\Delta_m(t)=\dot{\beta}(t)+\frac{\gamma_m}{2}\sin\varphi_m,\quad g(t)=\dot{\theta}(t)e^{-\beta_i(t)},\\
&\varphi(t)=\beta(t)+\frac{\pi}{2},\quad \gamma_m\cos\varphi_m=-2\dot{\beta}_i(t),
\end{aligned}
\end{equation}
by which $H_{\rm rot}(t)$ is simplified as
\begin{equation}\label{HamrotRedNon}
H_{\rm rot}(t)=\tilde{\Delta}_m(t)m^\dagger m+\frac{\tilde{\omega}_q(t)}{2}\sigma_z-\tilde{\Delta}_m(t)|\tilde{\alpha}(t)|^2.
\end{equation}
The Heisenberg equation~(\ref{DynInva}) is still capable to yield the dynamical invariant in the non-Hermitian case, when using the non-Hermitian rotated Hamiltonian $H_{\rm rot}(t)$. Under the inversion of $\tilde{\mathcal{V}}(t)$ in Eq.~(\ref{DispNon}), the evolution of dynamical invariant in the original picture is found to be
\begin{equation}\label{InvaspinOriNon}
\begin{aligned}
&\mu(t)=\tilde{\mathcal{V}}(t)\mu(0)\tilde{\mathcal{V}}^\dagger(t)\\
&=|-\tilde{\alpha}(t)\rangle\langle-\tilde{\alpha}(t)|\otimes|e\rangle\langle e|+|\tilde{\alpha}(t)\rangle\langle\tilde{\alpha}(t)|\otimes|g\rangle\langle g|.
\end{aligned}
\end{equation}
In contrast to the Hermitian result in Eq.~(\ref{InvaspinOri}), now the amplitude of the coherent state has an extra factor, i.e., $\alpha(t)\rightarrow\tilde{\alpha}(t)=\alpha(t)\exp[-\beta_i(t)]$, which is substantially attributed to the non-Hermitian component in Eqs.~(\ref{HamNon}) or (\ref{HamNonRed}) and reflects the probability nonconservation of the target state.

To save the probability conservation at the target moment $t=\tau$, the whole evolution can be parameterized to two stages as (i) $t\in[0,\tau/2]$ and (ii) $t\in[\tau/2,\tau]$. During one of them, the bosonic mode is subject to the loss effect and during the other, it is dominated by the gain effect. Due to Eqs.~(\ref{ConstrNon}) and (\ref{InvaspinOriNon}), at the turning point $t=\tau/2$, the switching from gain to loss or vice versa can be realized by a sudden change in $\beta_i(t)$. Despite the parameters might be non-differentiable functions of time, however, the system dynamics presented by the dynamical invariant is required to be continuous around the vicinity of the turning point. Thereby $\mu(t=\tau^-/2)\approx\mu(t=\tau^+/2)$ yields a criterion:
\begin{equation}\label{crite}
\left\langle-\tilde{\alpha}\left(\frac{\tau^-}{2}\right)\right|
-\tilde{\alpha}\left(\frac{\tau^+}{2}\right)\left\rangle
=\left\langle\tilde{\alpha}\left(\frac{\tau^-}{2}\right)\right|\tilde{\alpha}
\left(\frac{\tau^+}{2}\right)\right\rangle\approx1.
\end{equation}
According to Eq.~(\ref{ConstrNon}), the system can undergo a switching between the gain and loss effects by simply setting $\beta_i(t)$ in two stages as
\begin{equation}\label{critePra}
\begin{aligned}
\beta_i(t)&=\lambda_1\theta(t),\quad t\in[0,\tau/2],\\
\beta_i(t)&=\lambda_2\theta(t),\quad t\in[\tau/2,\tau],
\end{aligned}
\end{equation}
where the coefficients $\lambda_1$ and $\lambda_2$ are of opposite signs. Consequently, the criterion in Eq.~(\ref{crite}) can be simplified as $|\exp(-\lambda_1\theta(\tau^-/2))-\exp(-\lambda_2\theta(\tau^+/2))|^2\approx0$, which is always attainable under appropriate choices of $\lambda_1$ and $\lambda_2$. For the composite system initially in the state $|\psi(0)\rangle=|0\rangle\otimes|+\rangle$, it will evolve to a highly entangled state at $t=\tau$. The gain or loss rate of the qubit is set as $\gamma\tau=0.1$ with $\varphi_q=0$ when $t\in[0, \tau/2]$ and $\varphi_q=\pi$ when $t\in[\tau/2, \tau]$ to automatically normalize $|\psi(\tau)\rangle$.

\begin{figure}[htbp]
\centering
\includegraphics[width=0.8\linewidth]{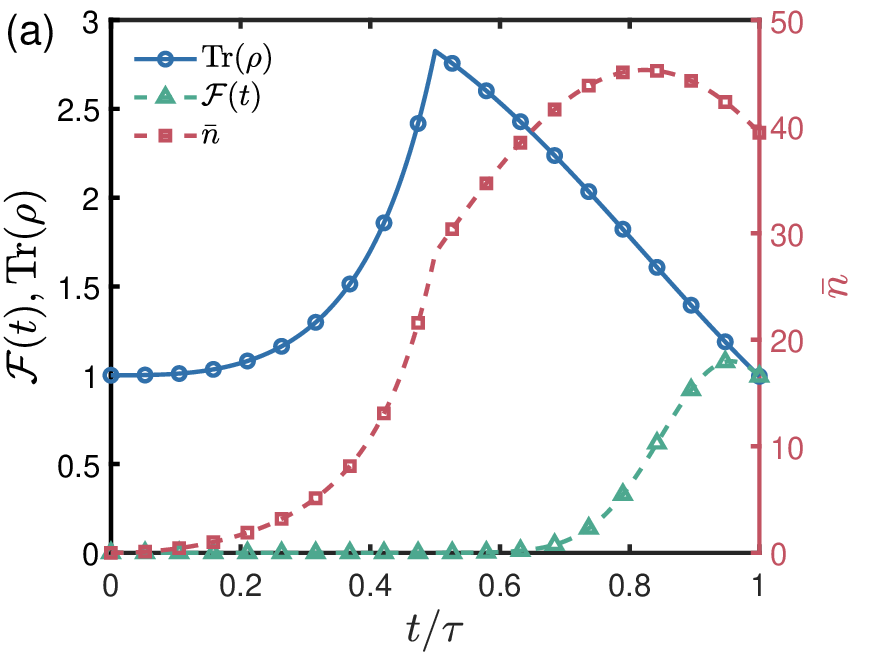}
\includegraphics[width=0.8\linewidth]{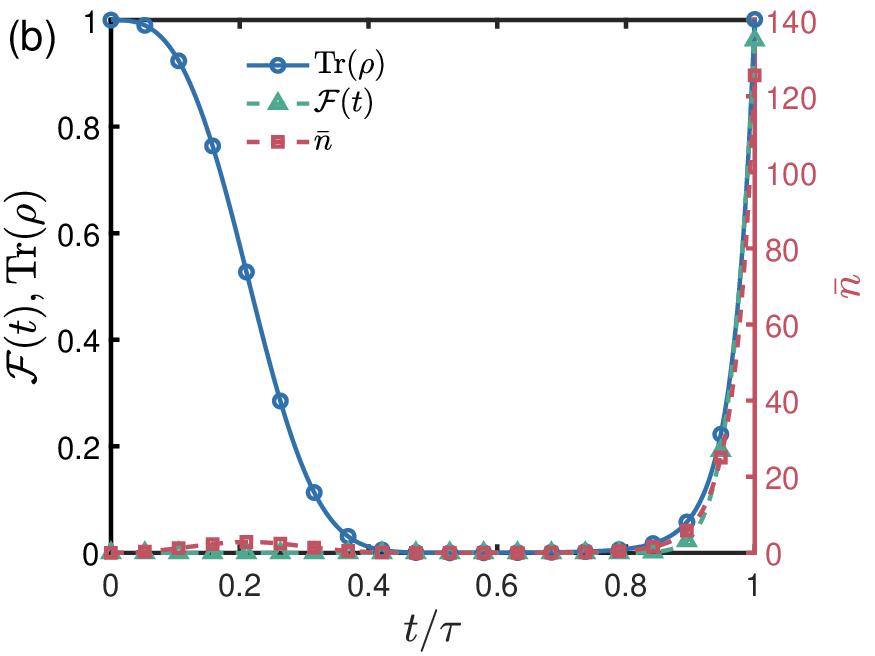}
\caption{Dynamics of the density-matrix trace about the whole system ${\rm Tr}[\rho(t)]$, the fidelity $\mathcal{F}(t)$ with respect to the target state $|\Psi(\tau)\rangle=(|\mathcal{C}_\alpha^+(\tau)\rangle\otimes|+\rangle
-|\mathcal{C}_\alpha^-(\tau)\rangle\otimes|-\rangle)/\sqrt{2}$, and the mean photon number $\bar{n}(t)$. The control period is set as $\tau\sim1\mu$s. The system is initially in the product state $|\psi(0)\rangle=|0\rangle\otimes|+\rangle$. The detuning of the bosonic mode $\Delta_m(t)$, the longitudinal coupling strength $g(t)$, the phase $\varphi(t)$, and the gain or loss rate of the bosonic mode $\gamma_m$ are set by Eq.~(\ref{ConstrNon}) where $\beta(t)=\pi t/(2\tau)+\pi/2$ and Eq.~(\ref{critePra}) about $\beta_i(t)$ with (a) $\lambda_1=1/20$, $\lambda_2=-0.143\lambda_1$, $\theta(t)=2\pi t/\tau$ and (b) $\lambda_1=-1/10$, $\lambda_2=0.142\lambda_1$, $\theta(t)=7\pi t/2\tau$. The coupling strength is larger than the gain or loss rates by one order in magnitude. $g/\gamma_m\sim10$ for $t\in[0, \tau/2]$ and $g/\gamma_m\sim20$ for $t\in[\tau/2, \tau]$. }\label{NonHermfide}
\end{figure}

Figure~\ref{NonHermfide} demonstrates the dynamics of the trace of the whole density matrix ${\rm Tr}[\rho(t)]$ with $\rho(t)=|\psi(t)\rangle\langle\psi(t)|$, the target state fidelity $\mathcal{F}(t)$, and the mean photon number $\bar{n}$. In Fig.~\ref{NonHermfide}(a), the system is subject to gain and loss effects in stages (i) and (ii), respectively, whereas in Fig.~\ref{NonHermfide}(b), the stages of gain and loss are reversed. In particular, in Fig.~\ref{NonHermfide}(a), the matrix trace first increases to be over unit in stage (i) and then gradually decreases to unit in stage (ii). And in Fig.~\ref{NonHermfide}(b), the behavior is roughly reversed. In both of them, ${\rm Tr}[\rho(t=\tau)]\approx1$ is attained. It is found that the target highly entangled states consisting of cat states with ultralarge amplitudes and qubit bases $|\pm\rangle$ can be obtained at the final moment. In Fig.~\ref{NonHermfide}(a), we have a nearly unit-fidelity target state $\mathcal{F}(\tau)\approx1$ with the mean photon number $\bar{n}\sim40$ when $t=\tau$; and in Fig.~\ref{NonHermfide}(b), we can generate a cat state with $\mathcal{F}(\tau)>0.962$ and $\bar{n}\sim120$. The final mean photon number is subject to the ratio of decoherence rate and the coupling strength between qubit and resonator and the control duration $\tau$, but again not relevant to the protocol itself.

\subsection{Assessment of protocol reliability}\label{fullmaster}

In this section, we first consider the system dynamics governed by the full master equation~(\ref{master}), retaining all the quantum jump terms associated with qubit, and then replace the abruptly changed parameters in Eq.~(\ref{critePra}) with smoothly varying parameters.

\begin{figure}[htbp]
\centering
\includegraphics[width=0.8\linewidth]{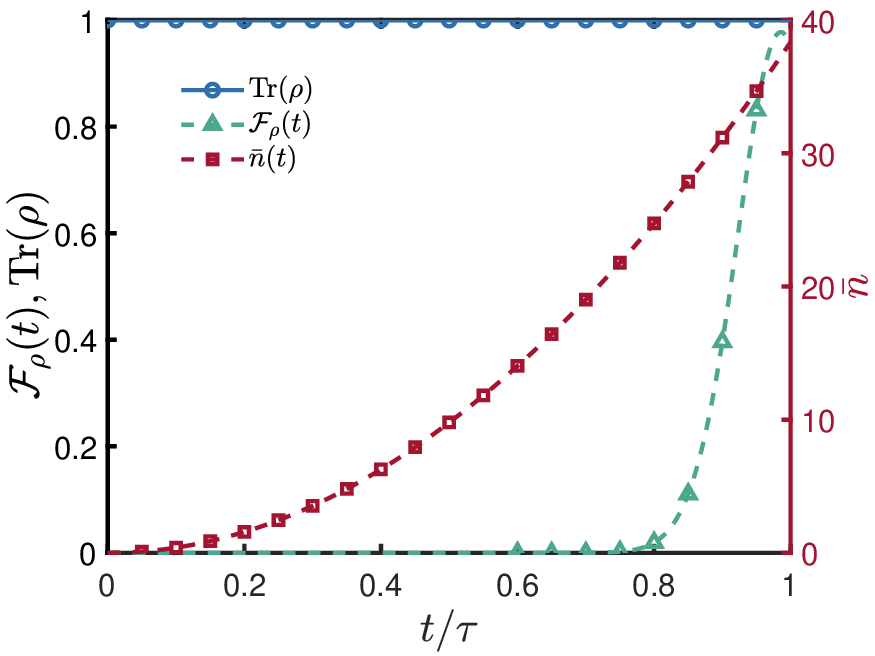}
\caption{Dynamics of the density-matrix trace about the whole system ${\rm Tr}[\rho(t)]$, the fidelity $\mathcal{F}_{\rho}(t)$ with respect to the target state $|\Psi(\tau)\rangle=(|\mathcal{C}_\alpha^+(\tau)\rangle\otimes|+\rangle
-|\mathcal{C}_\alpha^-(\tau)\rangle\otimes|-\rangle)/\sqrt{2}$, and the mean photon number $\bar{n}(t)$ under the full Lindblad master equation~(\ref{master}). The parameters are the same as Fig.~\ref{NonHermfide}(a).}\label{Commaster}
\end{figure}

The non-Hermitian dynamics in Sec.~\ref{opennojump} excludes the trajectories involving the quantum-jump terms for qubits in the master equation~(\ref{master}), which is equivalent to the postselection on the qubit state. One can assess the reliability of our UQC protocol formulated by the time-dependent Schr\"odinger equation under non-Hermitian Hamiltonian by comparing to the direct simulation using the full Lindblad master equation with the same set of control parameters. Figure~\ref{Commaster} is an analog to Fig.~\ref{NonHermfide}(a), that is used to generate a cat state of the same size $\bar{n}\approx40$, by holding all the quantum-jump terms. It is found that the trace is always maintained as unit and the final fidelity $\mathcal{F}_{\rho}(\tau)=0.951$ is slightly lower than unit in Fig.~\ref{NonHermfide}(a).

The success probability of postselection on qubit can be quantitatively estimated by $\mathcal{F}_\rho(\tau)=P_{nj}\mathcal{F}_\psi(\tau)+(1-P_{nj})\bar{\mathcal{F}}$~\cite{Minganti2020Hybrid}, where $P_{nj}$ denotes the no-jump probability associated with the non-Hermitian Hamiltonian~(\ref{HamNon}), $\mathcal{F}_\psi(\tau)\approx1$ is the target-state fidelity obtained by the non-Hermitian Schr\"odinger equation, and $\bar{\mathcal{F}}$ is the average fidelity of the trajectories involving quantum jumps. For an order of magnitude estimate, by assuming $\bar{\mathcal{F}}\sim1/2$ as the average fidelity of a random qubit state with respect to a pure target state, one can find a significant no-jump probability $P_{nj}\sim90.2\%$.

\begin{figure}[htbp]
\centering
\includegraphics[width=0.8\linewidth]{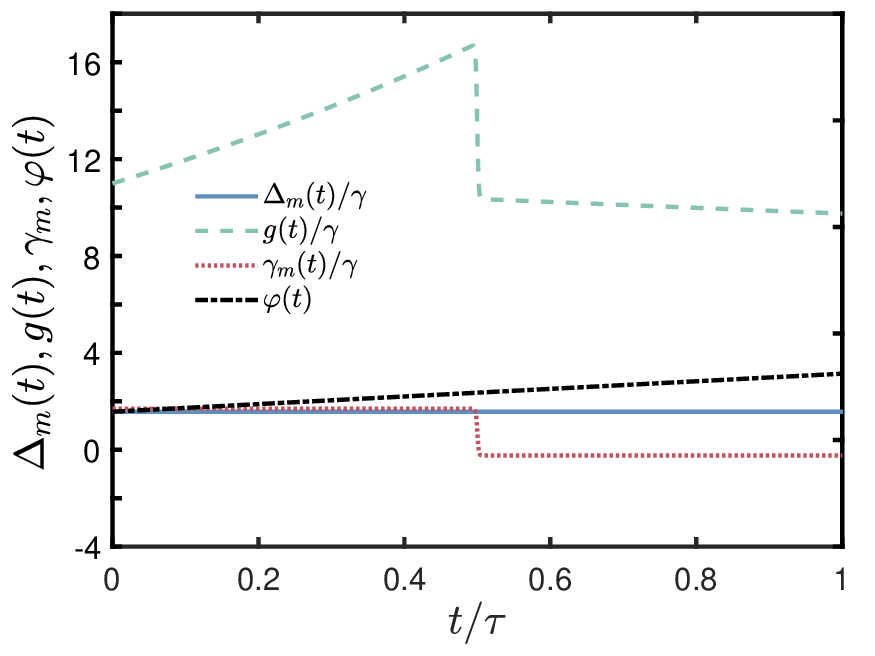}
\caption{Time-dependent profiles of the bosonic-mode detuning $\Delta_m(t)$, the longitudinal coupling strength $g(t)$, the gain or loss rate of the bosonic mode $\gamma_m$, and the coupling phase $\varphi(t)$ in Eq.~(\ref{ConstrNon}) used for plotting Fig.~\ref{NonHermfide}(b). Here $\beta_i(t)$ is given by Eq.~(\ref{paralogistic}) with $\eta=10^{-3}$.}\label{logistic}
\end{figure}

To account for the finite response time of realistic control devices, the abrupt switch between the gain and loss effects described by Eq.~(\ref{critePra}) can be replaced by a smooth transition using a logistic function over the control period $t\in[0,\tau]$ as
\begin{equation}\label{paralogistic}
\beta_i(t)=\left[\lambda_1+(\lambda_2-\lambda_1)
\frac{1}{1+e^{-(t-\tau/2)/\eta}}\right]\theta(t),
\end{equation}
where $\eta$ characterizes the switching timescale. Equation~(\ref{paralogistic}) reduces to Eq.~(\ref{critePra}) when $\eta\rightarrow0$. The relevant control parameters in Eq.~(\ref{ConstrNon}) then vary continuously between their gain- and loss-stage values as shown in Fig.~\ref{logistic}, thereby avoiding the sudden change at the turning point $t=\tau/2$. With these smooth control parameters, our protocol can still generate the same large-amplitude cat state with high fidelity as in Fig.~\ref{NonHermfide}(b).

In addition, the magnitude of the varying coupling strength $g(t)$ in Fig.~\ref{logistic} is about $g/\gamma_m\sim10$, which is accessible with the existing superconducting-circuit platform. The decay rate of the cavity is typically determined by the ratio of its eigenfrequency and quality factor. The latter is about $Q_c\sim10^3-10^8$ for a cavity frequency in the GHz regime~\cite{Reagor2016Quantum}, which means $\gamma_m$ is less than MHz. The relaxation and decoherence rates of superconducting qubit are also in the regime from sub-MHz to MHz. On the other hand, the longitudinal qubit-cavity interaction strength is related to the zero-point fluctuations of the superconducting resonator, which can be enhanced by increasing the characteristic impedance, enabling coupling strengths about $g/2\pi\sim10-10^3$ MHz~\cite{Didier2015Fast,Blais2021Circuit}. Thus the numerical simulation in Fig.~\ref{logistic} over the systematic parameters is feasible in experiments~\cite{Didier2015Fast,Blais2021Circuit}.

\section{Intrinsic cat states and four-legged cat states}\label{RealCat}

\subsection{Generation of intrinsic cat states}\label{SubReal}

In the context of the Schr\"odinger's gedanken experiment~\cite{Schrodinginger1935Die}, the cat state is intrinsically a highly entangled state of three components of the whole system, e.g., the dead or live cat, the two-level atom that in charge of the radioactive decay, and the broken or unbroken poison bottle. The latter two components can be modeled as two individual qubits. Then the intrinsic cat state can be defined as
\begin{equation}\label{IntCatDef}
|\mathcal{C}_{\rm int}\rangle=|-\alpha\rangle|e\rangle_1|e\rangle_2+|\alpha\rangle|g\rangle_1|g\rangle_2
\end{equation}
up to a normalization factor and $|e\rangle_j$ and $|g\rangle_j$, $j=1,2$, denote the excited and ground states of the $j$th qubit. To generate such a tripartite entangled state, we consider an open quantum system~\cite{Blais2004Cavity,
Billangeon2015Circuit,Didier2015Fast,Richer2016Circuit,Royer2017fasthighfidelity,
Harvey2018Coupling,Bottcher2022Parametric,Bosco2022Fully} with a Hamiltonian directly generalized from Eq.~(\ref{HamNon}) or (\ref{HamNonRed}):
\begin{equation}\label{Hamtwo}
\begin{aligned}
&H(t)=\tilde{\Delta}_m(t)m^\dagger m+\frac{\tilde{\omega}_q(t)}{2}\left[\sigma_z^{(1)}+\sigma_z^{(2)}\right]+g_1(t)\sigma_z^{(1)}\\
&\times\left(me^{i\varphi_1(t)}+{\rm H.c.}\right)
+g_2(t)\sigma_z^{(2)}\left(me^{i\varphi_2(t)}+{\rm H.c.}\right).
\end{aligned}
\end{equation}
Under the isotropic conditions that $g_1(t)=g_2(t)=g(t)/2$ and $\varphi_1(t)=\varphi_2(t)=\varphi(t)$, the system dynamics under the Hamiltonian~(\ref{Hamtwo}) can be restricted within the subspace spanned by $|ee\rangle\equiv|e\rangle_1\otimes|e\rangle_2$ and $|gg\rangle\equiv|g\rangle_1\otimes|g\rangle_2$. Then we have
\begin{equation}\label{HamtwoRed}
H(t)=H_e(t)\otimes|ee\rangle\langle ee|+H_g(t)\otimes|gg\rangle\langle gg|
\end{equation}
with the conditional Hamiltonian
\begin{equation}
\begin{aligned}
H_{e,g}(t)&=\tilde{\Delta}_m(t)m^\dagger m\pm\tilde{\omega}_q(t)\\
&\pm g(t)\left(me^{i\varphi(t)}+m^\dagger e^{-i\varphi(t)}\right).
\end{aligned}
\end{equation}
The dynamics of the bosonic mode is therefore conditioned on the two-qubit product states, i.e., $|ee\rangle$ and $|gg\rangle$, which is similar to Eq.~(\ref{HamNonRed}) conditioned by the single-qubit state.

In the rotating frame with respect to the conditional unitary transformation $\tilde{\mathcal{V}}(t)$, formulated by Eq.~(\ref{DispNon}) with the replacements of $|e\rangle\rightarrow|ee\rangle$ and $|g\rangle\rightarrow|gg\rangle$, the transformed Hamiltonian is expressed as
\begin{equation}\label{HamrotNonTwo}
\begin{aligned}
H_{\rm rot}(t)&=\tilde{\mathcal{V}}^\dagger(t)H(t)\tilde{\mathcal{V}}(t)
-i\tilde{\mathcal{V}}^\dagger(t)\frac{\partial\tilde{\mathcal{V}}(t)}{\partial t}\\
&=\left[\tilde{H}_e(t)+\frac{\tilde{\omega}_q(t)}{2}\right]\otimes|ee\rangle\langle ee|\\
&+\left[\tilde{H}_g(t)-\frac{\tilde{\omega}_q(t)}{2}\right]\otimes|gg\rangle\langle gg|.
\end{aligned}
\end{equation}
Here the conditional rotated Hamiltonian $\tilde H_e(t)$ and $\tilde H_g(t)$ share the same expression of their counterparts in Eq.~(\ref{HamrotNon}). Again, under the parametric constraints in Eq.~(\ref{ConstrNon}), the transformed Hamiltonian $H_{\rm rot}(t)$ can be reduced to
\begin{equation}\label{HamrotRedNonTwo}
\begin{aligned}
H_{\rm rot}(t)=&\tilde{\Delta}_m(t)(m^\dagger m-|\tilde{\alpha}(t)|^2)(|ee\rangle\langle ee|+|gg\rangle\langle gg|)\\+&\tilde{\omega}_q(t)(|ee\rangle\langle ee|-|gg\rangle\langle gg|),
\end{aligned}
\end{equation}
which satisfies the Heisenberg equation~(\ref{DynInva}) with the stationary dynamical invariant
\begin{equation}\label{InvaspinTwo}
\mu(0)=|0\rangle\langle0|\otimes|ee\rangle\langle ee|+|0\rangle\langle 0|\otimes|gg\rangle\langle gg|.
\end{equation}
By the inversion of $\tilde{\mathcal{V}}(t)$, the time evolution of $\mu(0)$ in the original picture is found to be similar to Eq.~(\ref{InvaspinOriNon}):
\begin{equation}\label{InvaspinOriNonTwo}
\begin{aligned}
&\mu(t)=\tilde{\mathcal{V}}(t)\mu(0)\tilde{\mathcal{V}}^\dagger(t)\\
&=|-\tilde{\alpha}(t)\rangle\langle-\tilde{\alpha}(t)|\otimes|ee\rangle\langle ee|+|\tilde{\alpha}(t)\rangle\langle\tilde{\alpha}(t)|\otimes|gg\rangle\langle gg|.
\end{aligned}
\end{equation}
Consequently, the criterion in Eq.~(\ref{crite}) still guarantees the continuity of the wave function at the turning point $\tau/2$. Starting from the initial state $|\Psi(0)\rangle=|0\rangle\otimes(|ee\rangle+|gg\rangle)/\sqrt{2}$, as consistent with the correlation between the radioactive atom and the poison bottle imagined by Schr\"odinger, the system will approach the intrinsic cat state $|\Psi(\tau)\rangle\approx(|-\alpha(\tau)\rangle|ee\rangle+|\alpha(\tau)\rangle|gg\rangle)/\sqrt{2}$. Since the generation procedure does not involve with projective measurement on qubit, one can verify that the intrinsic cat states can be generated with a close-to-unit fidelity and over $100$ photons when the parameters in Hamiltonian in Eq.~(\ref{HamtwoRed}) are of comparable magnitudes to those in Fig.~\ref{NonHermfide}(b).

\subsection{Generation of four-legged cat states}\label{fourcomponent}

Based on the hybrid qubit-resonator system governed by Eq.~(\ref{Hamtwo}), our protocol can be further extended to the generation of the large-amplitude four-legged cat states, i.e., $|\alpha(t)\rangle+|-\alpha(t)\rangle+|i\alpha(t)\rangle+|-i\alpha(t)\rangle$. To make the bosonic mode in a superposition of coherent states with four distinct phases, we assume that the coupling strengths are $g_1(t)=g_2(t)=g(t)/\sqrt{2}$ and the phases are $\varphi_1(t)=\varphi(t)+\pi/4$ and $\varphi_2(t)=\varphi(t)+3\pi/4$ with the phase $\varphi(t)$ subject to Eq.~(\ref{ConstrNon}). Then the non-Hermitian Hamiltonian~(\ref{Hamtwo}) is modified to
\begin{equation}\label{Hamfour}
\begin{aligned}
H(t)&=\tilde{\Delta}_m(t)m^\dagger m+\frac{\tilde{\omega}_q(t)}{2}\left[\sigma_z^{(1)}+\sigma_z^{(2)}\right]\\
&+ig(t)(|ee\rangle\langle ee|-|gg\rangle\langle gg|)\left(me^{i\varphi(t)}-m^\dagger e^{-i\varphi(t)}\right)\\
&+g(t)(|eg\rangle\langle eg|-|ge\rangle\langle ge|)\left(me^{i\varphi(t)}+m^\dagger e^{-i\varphi(t)}\right).
\end{aligned}
\end{equation}
Similar to Eq.~(\ref{DispNon}), the conditional unitary transformation $\tilde{\mathcal{V}}(t)$ for the generation of four-legged cat states takes the form of
\begin{equation}\label{DispNonfour}
\begin{aligned}
\tilde{\mathcal{V}}(t)&=D^\dagger[-\tilde{\alpha}(t)]\otimes|ee\rangle\langle ee|+D[\tilde{\alpha}(t)]\otimes|gg\rangle\langle gg|\\
&+D^\dagger[-i\tilde{\alpha}(t)]\otimes|eg\rangle\langle eg|+D[i\tilde{\alpha}(t)]\otimes|ge\rangle\langle ge|.
\end{aligned}
\end{equation}
In the rotating frame with respect to $\tilde{\mathcal{V}}(t)$, we have
\begin{equation}\label{HamrotNonfour}
\begin{aligned}
&H_{\rm rot}(t)=\tilde{\mathcal{V}}^\dagger(t)H(t)
\tilde{\mathcal{V}}(t)-i\tilde{\mathcal{V}}^\dagger(t)
\frac{\partial\tilde{\mathcal{V}}(t)}{\partial t}\\
&=\tilde{\mathcal{H}}_e(t)\otimes|ee\rangle\langle ee|+\tilde{\mathcal{H}}_g(t)\otimes|gg\rangle\langle gg|\\
&+\tilde{\mathcal{H}}_e^i(t)\otimes|eg\rangle\langle eg|+\tilde{\mathcal{H}}_g^i(t)\otimes|ge\rangle\langle ge|,
\end{aligned}
\end{equation}
where the conditional rotated Hamiltonian $\tilde{\mathcal{H}}_e(t)$ and $\tilde{\mathcal{H}}_g(t)$ are the same as those in Eq.~(\ref{HamrotNon}), and $\tilde{\mathcal{H}}_e^i(t)$ and $\tilde{\mathcal{H}}_g^i(t)$ take the similar forms as $\tilde{\mathcal{H}}_e(t)$ and $\tilde{\mathcal{H}}_g(t)$, respectively, with the replacement of $\tilde{\alpha}(t)\rightarrow i\tilde{\alpha}(t)$.

The linear terms of $H_{\rm rot}(t)$ in Eq.~(\ref{HamrotNonfour}) can be canceled by the constraints in Eq.~(\ref{ConstrNon}). And then we have
\begin{equation}\label{HamrotNonfourRed}
\begin{aligned}
H_{\rm rot}(t)&=\tilde{\Delta}_m(t)m^\dagger m+\frac{\tilde{\omega}_q(t)}{2}(|ee\rangle\langle ee|-|gg\rangle\langle gg|\\
&+|eg\rangle\langle eg|-|ge\rangle\langle ge|)-\tilde{\Delta}_m(t)|\tilde{\alpha}(t)|^2.
\end{aligned}
\end{equation}
Under this Hamiltonian, the dynamical invariant satisfying the Heisenberg equation~(\ref{DynInva}) is found to be
\begin{equation}\label{InvaspinTwoFour}
\begin{aligned}
\mu(0)&=|0\rangle\langle0|\otimes|ee\rangle\langle ee|+|0\rangle\langle 0|\otimes|gg\rangle\langle gg|\\
&+|0\rangle\langle0|\otimes|eg\rangle\langle eg|+|0\rangle\langle 0|\otimes|ge\rangle\langle ge|.
\end{aligned}
\end{equation}
And in the original picture it evolves as
\begin{equation}\label{InvaspinOriNonTwoFour}
\begin{aligned}
&\mu(t)=\tilde{\mathcal{V}}(t)\mu(0)\tilde{\mathcal{V}}^\dagger(t)
=|-\tilde{\alpha}(t)\rangle\langle-\tilde{\alpha}(t)|\otimes|ee\rangle\langle ee| \\ &+|\tilde{\alpha}(t)\rangle\langle\tilde{\alpha}(t)|\otimes|gg\rangle\langle gg|
+|-i\tilde{\alpha}(t)\rangle\langle-i\tilde{\alpha}(t)|\otimes|eg\rangle\langle eg|\\ &+|i\tilde{\alpha}(t)\rangle\langle i\tilde{\alpha}(t)|\otimes|ge\rangle\langle ge|
\end{aligned}
\end{equation}
under the inversion of $\tilde{\mathcal{V}}(t)$ in Eq.~(\ref{DispNonfour}).

Similar to the two-legged cat state generation, under the continuous condition for the dynamical invariant around the vicinity of the turning point between the gain and loss stages, i.e., $\mu(t=\tau^-/2)\approx\mu(t=\tau^+/2)$, Eq.~(\ref{InvaspinOriNonTwoFour}) gives rise to both Eq.~(\ref{crite}) and $\langle-i\tilde{\alpha}(\tau^-/2)|-i\tilde{\alpha}(\tau^+/2)\rangle=\langle i\tilde{\alpha}(\tau^-/2)|i\tilde{\alpha}(\tau^+/2)\rangle\approx1$. The two criteria are found to be identical to each other and they further guarantee the norm conservation of the wave-function of the composite system at the target moment $\tau$.

\begin{figure}[htbp]
\centering
\includegraphics[width=0.8\linewidth]{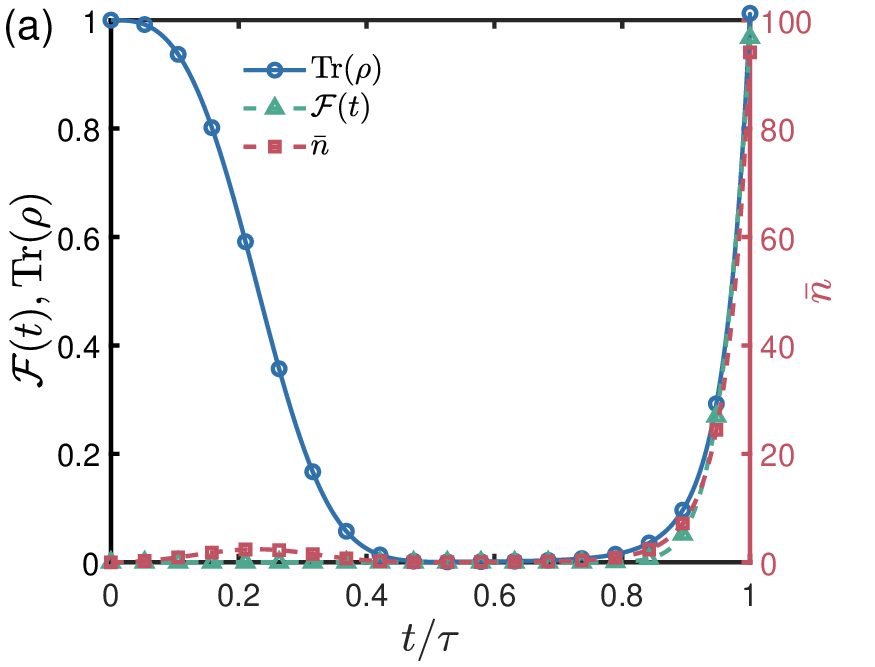}
\includegraphics[width=0.8\linewidth]{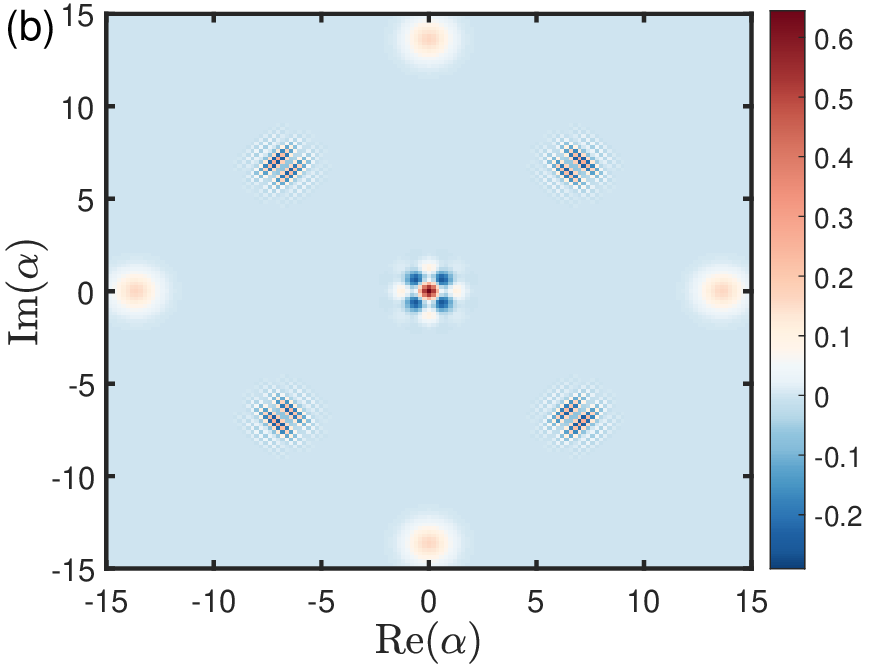}
\caption{(a) Dynamics of the density-matrix trace of ${\rm Tr}(\rho)$, the target state fidelity $\mathcal{F}(t)$ with respect to the final entangled state $|\Psi(\tau)\rangle$ in Eq.~(\ref{targetfour}), and the mean photon number $\bar{n}$.  (b) Wigner function for the relevant four-legged cat state $|\psi_b(\tau)\rangle=|\mathcal{C}_\alpha^+(\tau)\rangle
+|\mathcal{C}_{i\alpha}^+(\tau)\rangle$ at $t=\tau$, obtained by a projective measurement on $|+\rangle_1|+\rangle_2$ of the composite system. The parameters are set the same as Fig.~\ref{NonHermfide}(b) except $\theta(t)=3\pi t/\tau$. }\label{NonHermCatfour}
\end{figure}

By the dynamical invariant in Eq.~(\ref{InvaspinOriNonTwoFour}), the system initially in the product state $|\Psi(0)\rangle=|0\rangle|+\rangle_1|+\rangle_2$ will evolve to the highly entangled state as
\begin{equation}\label{targetfour}
\begin{aligned}
|\Psi(\tau)\rangle&=\frac{1}{2}\left[|\alpha\rangle|ee\rangle+|i\alpha\rangle|eg\rangle
+|-i\alpha\rangle|ge\rangle+|-\alpha\rangle|gg\rangle\right]\\
&=\frac{1}{2}\Big(|\mathcal{C}_{++}\rangle|+\rangle_1|+\rangle_2
+|\mathcal{C}_{-+}\rangle|-\rangle_1|+\rangle_2\\
&+|\mathcal{C}_{+-}\rangle|+\rangle_1|-\rangle_2
+|\mathcal{C}_{--}\rangle|-\rangle_1|-\rangle_2\Big),
\end{aligned}
\end{equation}
where the four-legged cat states are defined as
\begin{equation}\label{foucompCat}
\begin{aligned}
&|\mathcal{C}_{++}\rangle=|\mathcal{C}^+_\alpha(t)\rangle+|\mathcal{C}^+_{i\alpha}(t)\rangle,
\quad |\mathcal{C}_{-+}\rangle=|\mathcal{C}^-_\alpha(t)\rangle+|\mathcal{C}^-_{i\alpha}(t)\rangle,\\
&|\mathcal{C}_{+-}\rangle=|\mathcal{C}^-_\alpha(t)\rangle-|\mathcal{C}^-_{i\alpha}(t)\rangle,
\quad |\mathcal{C}_{--}\rangle=|\mathcal{C}^+_\alpha(t)\rangle-|\mathcal{C}^+_{i\alpha}(t)\rangle,
\end{aligned}
\end{equation}
with $|\mathcal{C}^\pm_{i\alpha}(t)\rangle=(|i\alpha(t)\rangle\pm|-i\alpha(t)\rangle)/\mathcal{N}$. For example, after performing the projective measurement on the basis $|+\rangle_1|+\rangle_2$ of two qubits, we have a four-legged bosonic cat state $|\mathcal{C}_{++}\rangle$ with a success probability of $25\%$.

Figure~\ref{NonHermCatfour}(a) presents the dynamics about the density-matrix trace ${\rm Tr}(\rho)$, the target state fidelity $\mathcal{F}(t)$ with respect to the highly entangled state $|\Psi(\tau)\rangle$ in Eq.~(\ref{targetfour}) with $|\alpha(\tau)|=3\pi$, and the mean photon number $\bar{n}$. It is found that when $t=\tau$, ${\rm Tr}(\rho)\approx1$, $\mathcal{F}(t)>0.965$, and $\bar{n}\sim95$. The Wigner function for the four-legged cat state $|\psi_b(\tau)\rangle=|\mathcal{C}_\alpha^+(\tau)\rangle+|\mathcal{C}_{i\alpha}^+(\tau)\rangle$ is demonstrated in Fig.~\ref{NonHermCatfour}(b), obtained by performing a projective measurement on the composite system. The four peaks and their mutual interference confirm the generated four coherent-state components.

\section{Conclusion and Discussion}\label{Conclud}

In summary, we establish a dynamical-invariant-based control protocol for the time-dependent hybrid qubit-resonator systems, under either Hermitian or non-Hermitian Hamiltonian, to deterministically generate large-amplitude and faithful two-legged cat states, intrinsic cat states, and four-legged cat states. The basic idea of our protocol is inherited from our universal quantum control theory, used to identify the unitary transformation associated with the coherent-state basis conditioned on the qubit state. Our protocol is economical in the operational complexity and resource overhead, since it involves only a single cavity mode longitudinally coupled to ancillary qubit(s) and requires neither manipulating nonlinear interactions nor multi-round control operations for cat-state generation. The model can be also indirectly implemented via displacement operations across various experimental platforms.

For open quantum systems, the non-Hermitian Hamiltonian can be derived from the Lindblad master equation under the postselection on qubit, and then UQC theory gives rise to the constraints associated with the gain or loss rates of the system, for activating the dynamical invariant. The controlled evolution can be divided into two stages to guarantee the probability conservation at the target moment. We confirmed that the effect of neglecting quantum-jump terms related to qubit is not significant. Essentially, this work extends our UQC theory to the hybrid discrete-continuous variable systems and paves an unconventional path toward generating the macroscopic quantum states via dynamical invariants.

Our protocol is based on the linear Hamiltonian~(\ref{Ham}) or (\ref{Hamprime}) that can be directly performed in circuit QED systems~\cite{Didier2015Fast,Blais2021Circuit}, while neglecting possible residual self-Kerr nonlinearity of the cavity, i.e., $H_K=K(m^\dagger m)^2$. Typically, the self-Kerr nonlinearity factor is about five orders of magnitude lower than the bosonic-mode eigenfrequency, e.g., $K/2\pi\sim48$ kHZ and $\omega_m/2\pi\sim5$ GHz~\cite{Didier2015Fast}. With a mean-field approximation, the frequency shift induced by the Kerr term is about $\delta\omega_m/\omega_m=K\langle m^\dagger m\rangle/\omega_m\sim10^{-3}$ for the large-amplitude cat states with $\langle m^\dagger m\rangle=\bar{n}\sim100$. It indicates that the self-Kerr nonlinearity has a negligible effect on the linear dynamics in this work.

\section*{Acknowledgments}

We acknowledge grant support from the National Natural Science Foundation of China (Grant No. U25A20199) and the ``Pioneer'' and ``Leading Goose'' R\&D of Zhejiang Province (Grant No. 2025C01028).

\appendix

\section{Derivation from Lindblad master equation to non-Hermitian Hamiltonian}\label{EffHam}

This appendix provides a detailed derivation of the non-Hermitian Hamiltonian~(\ref{HamNon}) from the Lindblad master equation for a hybrid discrete-continuous variable system. While the quantum-jump terms for qubit are neglected after postselection~\cite{Han2024Measuring}, those for bosonic mode are fully retained~\cite{Metelmann2015Nonreciprocal,Wang2019Nonreciprocity}. In general, the dynamics of an open system can be described by
\begin{equation}\label{master}
\frac{d}{dt}\rho=-i[H_{\rm coh}, \rho]+\gamma\mathcal{L}[\sigma_-]\rho+\gamma_m\mathcal{L}[m]\rho,
\end{equation}
where $H_{\rm coh}=\Delta_mm^\dagger m+(\omega_q/2)\sigma_z+g\sigma_z[m\exp(i\varphi)+m^\dagger\exp(-i\varphi)]$ represents the eigen-energies of the qubit and the bosonic mode and the longitudinal interaction between them. The Lindblad superoperators are defined as $\mathcal{L}[o]\rho=o\rho o^\dagger-\{o^\dagger o,\rho\}/2$ with $o=\sigma_-,m$. The first and the second superoperators are associated with the individual dissipations of the qubit and the bosonic mode, with the decay rates $\gamma$ and $\gamma_m$, respectively.

Under Eq.~(\ref{master}), an arbitrary Schr\"odinger-picture operator $\mathcal{O}_S$ can be connected with the Heisenberg-picture operator $\mathcal{O}_H(t)$ through
\begin{equation}\label{masterSH}
\begin{aligned}
&{\rm Tr}\left[\mathcal{O}_S\dot{\rho}(t)\right]\\
=&{\rm Tr}\left[\mathcal{O}_S\left(-i[H_{\rm coh},\rho]+\gamma\mathcal{L}[\sigma_-]\rho+\gamma_m\mathcal{L}[m]\rho\right)\right]\\
=&{\rm Tr}\left[\left(i[H_{\rm coh},\mathcal{O}_S]+\gamma\mathcal{L}^\dagger[\sigma_-]\mathcal{O}_S
+\gamma_m\mathcal{L}^\dagger[m]\mathcal{O}_S\right)\rho\right]\\
=&{\rm Tr}\left[\dot{\mathcal{O}}_H(t)\rho(0)\right],
\end{aligned}
\end{equation}
where the Hermitian conjugate superoperator is defined as $\mathcal{L}^\dagger[o]\mathcal{O}_S\equiv o^\dagger\mathcal{O}_So-\{o^\dagger o,\mathcal{O}_S\}/2$. The derivation from the second line to the third one in Eq.~(\ref{masterSH}) has used the cyclic property of the trace. Consequently, the time derivative of $\mathcal{O}_H(t)$ can be expressed by the adjoint Lindblad master equation as
\begin{equation}\label{OHdynamic}
\frac{d}{dt}\mathcal{O}_H(t)=i[H_{\rm coh},\mathcal{O}_S]+\gamma\mathcal{L}^\dagger[\sigma_-]\mathcal{O}_S
+\gamma_m\mathcal{L}^\dagger[m]\mathcal{O}_S.
\end{equation}

Using Eq.~(\ref{OHdynamic}) and neglecting the quantum jump terms about the qubit, the dynamics of the operators $\sigma_-$ and $m$ can be obtained as
\begin{equation}\label{abdynamic}
\begin{aligned}
\frac{d}{dt}\sigma_-(t)&=-i\omega_q\sigma_--\frac{1}{2}\gamma\sigma_-
-i2g\sigma_-\left(me^{i\varphi}+m^\dagger e^{-i\varphi}\right),\\
\frac{d}{dt}m(t)&=-i\Delta_mm-\frac{\gamma_m}2m-ige^{-i\varphi}\sigma_z.
\end{aligned}
\end{equation}
The system dynamics described by Eq.~(\ref{abdynamic}) is equivalent to that governed by the Hamiltonian~(\ref{HamNon}) with $\varphi_m=\varphi_e-\pi=\varphi_g=0$ and $\gamma_e=\gamma_q=\gamma$.

\bibliographystyle{apsrevlong}
\bibliography{ref}

\end{document}